\documentclass[preprint,showpacs,showkeys,aps,prb]{revtex4}
\usepackage[english]{babel}
\usepackage[dvips]{graphicx}

\begin{document}

\title{
Deterministic Time-Reversible Thermostats : \\
Chaos, Ergodicity, and the Zeroth Law of Thermodynamics
} 

\author{
{\bf William Graham Hoover}, Ruby Valley Research Institute \\
Highway Contract 60 Box 601, Ruby Valley Nevada USA 89833 ; \\
\vspace{2 mm}
{\bf Julien Clinton Sprott}, Department of Physics \\
University of Wisconsin Madison, Wisconsin 53706 ;\\
\vspace{2 mm}
{\bf Puneet Kumar Patra}, Advanced Technology Development Center \\
Department of Civil Engineering, Indian Institute of Technology \\
Kharagpur, West Bengal, India 721302 . \\
}

\date{\today}



\vspace{0.1cm}

\begin{abstract}
The relative stability and ergodicity of deterministic time-reversible thermostats, both
singly and in coupled {\it pairs}, are assessed through their Lyapunov spectra. Five types
of thermostat are coupled to one another through a single Hooke's-Law harmonic spring.
The resulting dynamics shows that three specific thermostat types,  Hoover-Holian, Ju-Bulgac,
and Martyna-Klein-Tuckerman, have very similar Lyapunov spectra in their equilibrium
four-dimensional phase spaces and when coupled in equilibrium or nonequilibrium pairs. All
three of these oscillator-based thermostats are shown to be {\it ergodic}, with smooth
analytic Gaussian distributions in their extended phase spaces ( coordinate, momentum, and
two control variables ).  Evidently these three ergodic and time-reversible thermostat types
are particularly useful as statistical-mechanical thermometers and thermostats. Each of them
generates Gibbs' universal canonical distribution internally as well as for systems to which
they are coupled.  Thus they obey the Zeroth Law of Thermodynamics, as a good heat bath should.
They also provide dissipative heat flow with relatively small nonlinearity when two or more
such bath temperatures interact and provide useful deterministic replacements for the
stochastic Langevin equation.
\end{abstract}

\maketitle

\section{Deterministic Time-Reversible Thermostats and Ergodicity}

In the early days of equilibrium many-body simulation, intercomparisons of results from
constant-temperature Monte Carlo with those from constant-energy molecular dynamics were 
indirect, requiring the conversion of isothermal $NVT$ data to isoenergetic $NVE$ data.
Only for hard disks and hard spheres, where temperature and energy are proportional, are
the two ensembles identical\cite{b1}.

Shuichi Nos\'e provided a conceptual breakthrough linking temperature $T$ and energy $E$
by discovering an isothermal canonical-ensemble molecular dynamics\cite{b2,b3}.  He started
out by including an additional time-reversible ``time-scaling variable'' $s$ in his novel
Hamiltonian :
$$
{\cal H}_{\rm Nos\acute{e}} \equiv [ \ K(p)/s^2 \ ] + \Phi(q) +
(p_s^2/2M) + (\#kT/2)\ln(s^2) \ .
$$
Here $\#$ is the number of degrees of freedom, {\it including} $s$ , and $p_s$ is the
momentum conjugate to $s$ . $M$ is a free parameter.  Getting to the canonical ensemble
from Nos\'e's Hamiltonian equations of motion requires just two more steps ; [ i ]
``scaling the time'', multiplying all of Nos\'e's time derivatives by $s$ :
$$
\{ \ \dot q \rightarrow s\dot q \ ; \ \dot p \rightarrow s\dot p \ \} \ ; \
\dot s \rightarrow s\dot s \ ; \ \dot p_s \rightarrow s\dot p_s \ ;
$$
then [ ii ] replacing all the scaled momenta $\{ \ (p/s) \ \}$ by $\{ \ p \ \}$ .  Nos\'e
showed that the resulting distribution for the $\{ \ q,p \ \}$ is Gibbs' canonical
distribution.

Starting instead with the ``Nos\'e-Hoover'' equations of motion\cite{b4,b5,b6,b7} ,
$$                                                                                               
\{ \ \dot q = (p/m) \ ; \ \dot p = F(q) - \zeta p \ \} \ ; \ \dot \zeta =                  
\sum^{\#=ND} [ \ (p^2/mkT) - 1 \ ]/\tau^2 \ ,                                                      
$$
Hoover showed that Gibbs' canonical distribution, with $s$ absent and with $\# = ND$
{\it not} including that extraneous $s$ variable, is a stationary solution of the extended
phase-space continuity equation ,
$$
(\partial f/\partial t) = 0 = -\sum^\# (\partial f\dot q/\partial q)
 - \sum^\# (\partial f\dot p/\partial p) - (\partial f\dot \zeta/\partial \zeta) \ .
$$
Here $\zeta$ is a ``friction coefficient'' proportional to Nos\'e's $p_s$ and the free 
parameter $\tau$ takes the place of Nos\'e's parameter $M$ .  The stationary distribution
function for the Nos\'e-Hoover motion equations is canonical in the $\{ \ q,p \ \}$ and
Gaussian in the friction coefficient $\zeta$ :
$$
f(q,p,\zeta) \propto e^{-{\cal H}/kT}e^{-\tau^2\zeta^2/2} \ ; \
{\cal H} = \Phi(q) + K(p) \ .
$$
The relaxation time $\tau$ controls the decay rate of velocity fluctuations in a $D$-dimensional
$N$-body Hamiltonian system.

In order for averages using the Nos\'e-Hoover equations to agree with canonical-ensemble
averages, it is necessary in principle that the dynamics be ``ergodic'', meaning that it
must  reach all of the $\{ \ q,p \ \}$ phase-space states.  Of course in practice only a
representative sampling of such states can be achieved.  Whether or not the motion equations
are ``ergodic'' in this sense, {\it capable} of reaching all of the states described by Gibbs'
canonical ensemble, depends upon the details of the underlying Hamiltonian ${\cal H}(q,p)$ .
For a simple harmonic oscillator Hoover pointed out that neither the four original unscaled
Nos\'e equations nor the three Nos\'e-Hoover motion equations are ergodic :
$$
\dot q = +(p/s^2) \ ; \ \dot p = - q \ ; \
M\ddot s = \dot p_s = [ \ (p^2/s^3) - (2/s) \ ] \ ; \ [ \ {\rm Nos\acute{e}, \ Not \ Ergodic} \ ] \ .
$$
$$
\dot q = +p \ ; \ \dot p = - q - \zeta p \ ; \ \dot \zeta = [ \ p^2 - 1 \ ]/\tau^2 \ ; \ 
[ \ {\rm Nos\acute{e}-Hoover, \ Not \ Ergodic} \ ] \ .
$$
These Nos\'e equations are singular while the Nos\'e-Hoover equations are not.

With the relaxation time $\tau$ equal to unity, numerical Nos\'e-Hoover calculations
show that about five percent of the initial conditions drawn from the Gaussian
distribution $e^{-q^2/2}e^{-p^2/2}e^{-\zeta^2/2}$ are chaotic, making up a ``chaotic
sea'' in the $(q,p,\zeta)$ space which is penetrated by an infinite number of holes.
The remaining 95\% of initial conditions generate regular periodic toroidal solutions
which fill in these holes.  Evidently this Nos\'e-Hoover thermostated oscillator is far
from ``ergodic'' [ where in this case an ergodic solution would have a smooth analytic
Gaussian density throughout $(q,p,\zeta)$ space ] :
$$
f_{\rm Gibbs}(q,p,\zeta) \propto e^{(-q^2/2)}e^{(-p^2/2)}e^{(-\zeta^2/2)} \
[ \ {\rm Ergodic} \ ] \ .
$$

In 1990 Bulgac and Kusnezov reiterated the usefulness of the phase-space continuity equation in
formulating more complicated motion equations ( with two or three thermostating control variables ).
Shortly thereafter they were able to use this approach to fill out the complete Gibbs' distribution for
prototypical Hamiltonian systems like the harmonic oscillator and the two-well ``Mexican Hat''
problems\cite{b8,b9,b10}.

There is no {\it foolproof} test for ergodicity.  The only reliable diagnostic for space-filling
ergodicity is the Lyapunov spectrum\cite{b11,b12,b13,b14,b15,b16,b17,b18,b19}.  If this spectrum,
which measures the long-time-averaged sensitivity of the dynamics to initial conditions, is not
only chaotic, but also the {\it same} for {\it all} initial conditions, the system is likely 
ergodic.  For
the harmonic oscillator this means that {\it all} $(q,p)$ oscillator states, all the way to 
infinity,  will {\it eventually}
occur.  In nonergodic systems the long-time-averaged spectrum depends upon the initial conditions.
In ergodic systems the long-time spectrum is always the same.  By 1996 there were three known
simple types of differential equations [ HH = Hoover-Holian, JB = Ju-Bulgac, MKT =
Martyna-Klein-Tuckermann ] which were thought to produce ergodicity in a one-dimensional harmonic
oscillator\cite{b9,b17,b18,b19}. In addition to propagating the phase variables ($q,p$) all three
included {\it two} additional control variables ($\zeta,\xi$) capable of generating Gibbs' canonical
distribution.  For the simple harmonic oscillator these three models take the form :
$$
\{ \ \dot q = p \ ; \ \dot p = -q - \zeta p -\xi p^3 \ ; \ \dot \zeta = p^2 - T \ ; \
\dot \xi = p^4 - 3Tp^2 \ \} \ [ \ {\rm HH} \ ] \ ;
$$
$$
\{ \ \dot q = p \ ; \ \dot p = -q - \zeta ^3p -\xi p^3 \ ; \ \dot \zeta = p^2 - T \ ; \ 
\dot \xi = p^4 - 3Tp^2 \ \} \ [ \ {\rm JB} \ ] \ ;
$$
$$
\{ \ \dot q = p \ ; \ \dot p = -q - \zeta p \ ; \ \dot \zeta = p^2 - T - \xi \zeta\ ; \ 
\dot \xi = \zeta^2 - T \ \} \ [ \ {\rm MKT} \ ] \ .
$$
So far no one-thermostat oscillator system ( with just a single control variable ) has been shown
to be ergodic though it certainly is possible that such a one will be discovered.  We explicitly
include the temperature $T$ in
all of these models in order to set the stage for {\it nonequilibrium} systems incorporating both
``cold'' and ``hot'' degrees of freedom.  It is worth pointing out that not all two-thermostat
oscillator systems are ergodic.  Patra and Bhattacharya showed that their very reasonable set
of doubly-thermostated oscillator equations is not ergodic\cite{b14,b19} :
$$
\{ \ \dot q = p - \xi q\ ; \ \dot p = -q - \zeta p \ ; \ \dot \zeta = p^2 - T \ ; \ 
\dot \xi = q^2 - T \ \} \ [ \ {\rm PB} \ ] \ .
$$

Our work here has three different aspects.  First we explore the equilibrium Lyapunov instability
which facilitates the ergodicity of all three thermostats.  In the equilibrium case,
where the phase-space distribution is a smooth Gaussian, the Lyapunov instability
embedded in that Gaussian has a totally different multifractal character.  We show that
the  equilibrium HH, JB, and MKT thermostats obey the Zeroth Law when coupled with one another
or with other Hamiltonian systems.  These thermostats can  provide Gibbs' complete canonical
distribution provided that any internal energy barriers are not too large relative to $kT$.
Second we consider {\it nonequilibrium} cases, where heat flows between thermostats with the
thermostat temperatures set at different values.  The nonequilibrium dynamics is still ergodic but
the phase-space distribution is no longer a smooth Gaussian.  It is instead an intimate multifractal
combination of the attractor-repellor pairs common to macroscopic time-reversible, but
dissipative, systems.  We also illustrate the application of ergodic thermostats to kinetic
barrier-crossing problems. Last of all, we summarize the lessons learned in this study of three
different sorts of applications of our ergodic time-reversible thermostats.

\section{Lyapunov Exponents--Thermostated Harmonic Oscillators}

The most convincing evidence for the lack of ergodicity with one thermostat variable, and its
lack or presence in two-thermostat systems, is the Lyapunov spectrum. Here we concentrate on
thermostated oscillators, the prototypical ``difficult'' case.  A harmonic-oscillator system of
four ordinary differential equations, such as the HH, JB, MKT, and PB sets for
$(\dot q,\dot p,\dot \zeta, \dot \xi)$ has four such exponents
$(\lambda_1,\lambda_2,\lambda_3,\lambda_4)$ . These exponents describe the deformation of an
infinitesimal hypersphere or hypercube in the four-dimensional phase space which contains the
motion.  Shimada and Nakashima\cite{b11}, as well as Benettin's group\cite{b12} described
general approaches to determining the spectrum of exponents.

\begin{figure}
\vspace{1 cm}
\includegraphics{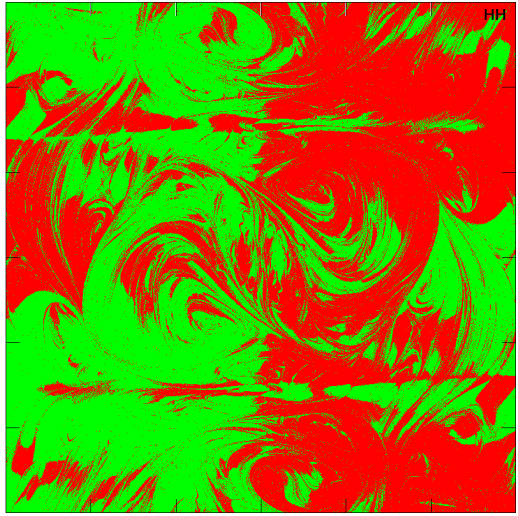}
\caption{
The sign of the local (time-dependent) value of the largest Lyapunov exponent for the
Hoover-Holian oscillator is shown in the $(q,p,0,0)$ plane.  Because a mirror reflection of the
dynamics ( perpendicular to the $q$ axis ) changes the signs of both $q$ and $p$ the Figure reveals
an inversion symmetry.  In the upper half plane red indicates a positive exponent and green a
negative.  The colors are reversed in the lower half plane.  The probability density in the
cross-sectional plane is a simple Gaussian in $q$ and $p$ .  Nevertheless the local Lyapunov
exponents vary in a multifractal manner throughout the four-dimensional space, reflecting their
sensitivity to bifurcations in their past histories.  In Figures 1-4 both $q$ and $p$ range
from -3 to +3 .
}
\end{figure}

In an $n$-dimensional space their algorithms require the solution of $n+1$ sets of $n$
equations.  In addition to a ``reference trajectory'' the equations describe the motion
of an associated orthonormal  set of $n$ comoving basis vectors, centered on the reference
trajectory and locating $n$ nearby ``satellite trajectories'' in the $n$-dimensional space.
The first Lyapunov exponent gives the time-averaged rate at which two neighboring solutions
of the equations diverge from one another,
$\langle \ \dot \delta(t)\simeq e^{\lambda_1t} \ \rangle $ .  The rate at which the {\it area}
defined by three nearby solutions ( the reference and two others ) diverges ( or converges )
$\simeq e^{\lambda_1t}e^{\lambda_2t}$ defines $\lambda_2$, while the rate at which the volume
defined by four solutions and the hypervolume defined by all five of them define $\lambda_3$
and $\lambda_4$ . Although these exponents are typically ``paired'' for Hamiltonian systems,
with
$$
[ \ \lambda_1 \equiv \langle \ \lambda_1(t) \ \rangle = -\langle\ \lambda_4(t) \ \rangle \ ; \ 
\lambda_2 \equiv \langle\ \lambda_2(t) \ \rangle =-\langle\ \lambda_3(t) \ \rangle = 0 \ ]    
\rightarrow \sum_1^4\lambda_i \equiv 0 \ ,                                                                    $$
with the sum total equal to zero, as implied by Liouville's Theorem, none of the two-thermostat
systems is Hamiltonian so that this instantaneous symmetry is missing in the time-dependent
exponents $\{ \ \lambda(t) \ \}$ . But because the equations of motion are time-reversible the
time-averaged spectra $\{ \ \lambda \ \}$ {\it are} symmetric.

For an {\it ergodic} system the long-time-averaged spectrum of exponents should be
{\it independent} of the initial conditions.  In practice, for the harmonic-oscillator system,
a few hundred oscillations is time enough to indicate whether or not a chosen initial condition
( either from a grid or chosen randomly ) converges to a spectrum close to the unique
long-time-averaged Lyapunov spectrum.  Estimates of these long-time-averaged exponents, after
a {\it time} $t = 40 \ 000 \ 000$ using a fourth-order Runge-Kutta integrator with an adaptive
timestep, are as follows :
$$                                                                                                            \{ \ \lambda \ \}_{\rm HH} = +0.068_0, \ +0.000, \ -0.000, \ -0.068_0 \ ;                                     $$
in the Hoover-Holian case, and
$$
\{ \ \lambda \ \}_{\rm JB} = +0.079_7, \ +0.0000, \ -0.0000, \ -0.079_7 \ ;
$$ 
in the Ju-Bulgac case, and  
$$
\{ \ \lambda \ \}_{\rm MKT} = +0.066_5, \ +0.0000, \ -0.0000, \ -0.066_5 \ .
$$
in the Martyna-Klein-Tuckerman case.

\begin{figure}
\vspace{1 cm}
\includegraphics{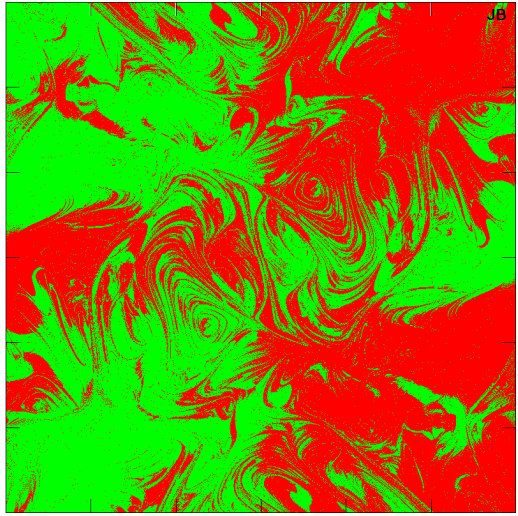}
\caption{
In the upper half plane red indicates a positive local Lyapunov exponent for the ergodic Ju-Bulgac
oscillator model.  The inversion symmetry ( with change of color ) is obeyed by all these
time-reversible oscillator models.  The probability density in the four-dimensional space is
Gaussian in $q$, $p$, $\zeta$, and $\xi^2$ as it is also in the $(q,p,0,0)$ plane displayed
here.
}
\end{figure}

The increase in the largest Lyapunov exponent with the number of quartic forces ( none for MKT,
one for HH, and two for JB ) suggests, as emphasized by Bulgac and Kusnezov\cite{b9}, that these
terms are particularly well-suited to promoting chaos and/or ergodicity.  It seems likely that
sextic forces, controlling $ \langle \ p^6 \ \rangle $ , would have no particular advantages and
would slow numerical work.

The probability densities for the three cases follow easily from the phase-space continuity
equation.  This makes it is easy to check for ergodicity by comparing relatively short-time
estimates for the Lyapunov spectrum starting out with initial conditions chosen according to
the stationary multivariable Gaussian distributions.

The ergodicity of the Martyna-Klein-Tuckerman oscillator was called into question by Patra
and Bhattacharya in 2014\cite{b14,b15,b16}, based on apparent ``holes'' in a $(q,p,\zeta,\xi)$
= $(q,p,-1,+ 1)$ double cross-section plane, the analog of a Poincar\'e
plane for a three-dimensional flow problem.  To resolve this question, the subject of the
2014 Ian Snook Prize\cite{b15}, we chose one million different initial conditions from the
appropriate four-dimensional Gaussian distributions for the HH, JB, and MKT cases and
determined that {\it all of them} were fully consistent with a unique chaotic spectrum.
Each of these initial conditions was followed for one million fourth-order Runge-Kutta
timesteps of 0.005 each.

The alternative possibility, a regular ( nonchaotic ) solution with all four of the long-time-averaged
Lyapunov exponents equal to zero, is easy to distinguish from the chaotic case.  As an additional
check the velocity moments generated by these three ergodic thermostat types likewise reproduce Gibbs'
values for the second, fourth, and sixth velocity moments with five-figure accuracy :
$$
\langle \ p^{\{ \ 2,4,6 \ \}} \ \rangle = \{ \ 1.00000, \ 3.0000, \ 15.000 \ \} \ .
$$

By contrast, the motion equations for a Nos\'e-Hoover harmonic oscillator with a single friction 
coefficient $\zeta(t)$ and unit relaxation time ,
$$
\{ \ \dot q = p \ ; \ \dot p = - q - \zeta p \ ; \ \dot \zeta = p^2 - 1 \ \} \ ,
$$
are known {\it not} to be ergodic despite their simple three-dimensional Gaussian solution of the
phase-space continuity equation. Choosing initial conditions from this stationary distribution
$f_{\rm NH} \propto e^{-(1/2)(q^2+p^2+\zeta^2)}$
and measuring the tendency of a nearby initial condition to separate gives the largest {\it Lyapunov
exponent} for the chosen initial condition, $\lambda_1 = \langle \ (d/dt)(\ln(\delta r)) \ \rangle$ .
Because this system is conservative and time-reversible the long-time-averaged Lyapunov spectrum is
symmetric, $\{ \ +\lambda_1,0,-\lambda_1 \ \}$ , and sums to zero.
$10 \ 000$ initial conditions followed for a time $t = 50 \ 000$ divide neatly into two groups :
$$
0.00002 < \lambda_1 < 0.00014 \ ; \ 0.006 < \lambda_1 < 0.017 \ .
$$
557 initial conditions correspond to a Lyapunov exponent significantly different to zero while
9443 correspond to nonchaotic toroidal solutions for which all three Lyapunov exponents vanish.
Evidently roughly 6\% of the stationary measure is chaotic ( with a positive time-averaged largest
Lyapunov exponent ).  The remaining 94\% consists of regular tori with a vanishing Lyapunov
spectrum :
$$
\langle \ \lambda_1,\lambda_2,\lambda_3 \ \rangle = \{ \ 0,0,0 \ \} \ .
$$

\begin{figure}
\vspace{1 cm}
\includegraphics{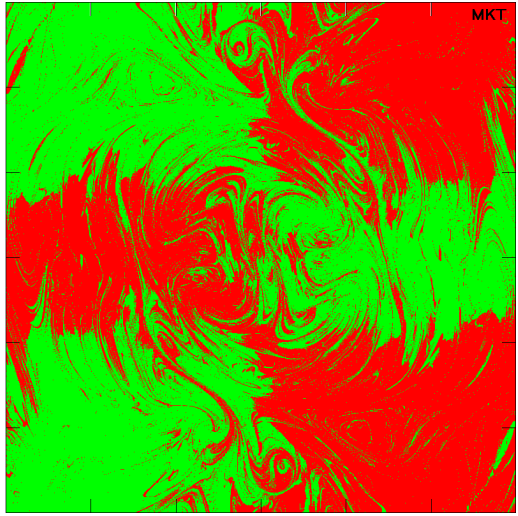}
\caption{
Red indicates a positive Lyapunov exponent in the upper half plane, negative in the lower.  If
the equations of motion are run backward ( by changing the righthandsides of all four equations )
the effect is to reflect the section shown here, changing the sign of $q$.  Like the other models
this Martyna-Klein-Tuckerman oscillator has a four-dimensional Gaussian distribution in its
$(q,p,\zeta,\xi)$ phase space which the dynamics samples ergodically.
}
\end{figure}

In the language of control theory the Nos\'e-Hoover equations control the time-averaged value
of the kinetic temperature, $\langle \ (p^2/mk) \ \rangle \equiv T$ .  Subsequent work
strongly suggests that an ergodic oscillator requires control of at least two moments, not just
one.  Such an approach requires four or more ordinary differential equations. Bulgac and
Kusnezov\cite{b9} as well as Ju and Bulgac\cite{b10} added a fifth equation so as to be able to
thermostat a free particle ( or a rotating and translating cluster of particles ) undergoing
Brownian motion.

\begin{figure}
\vspace{1 cm}
\includegraphics{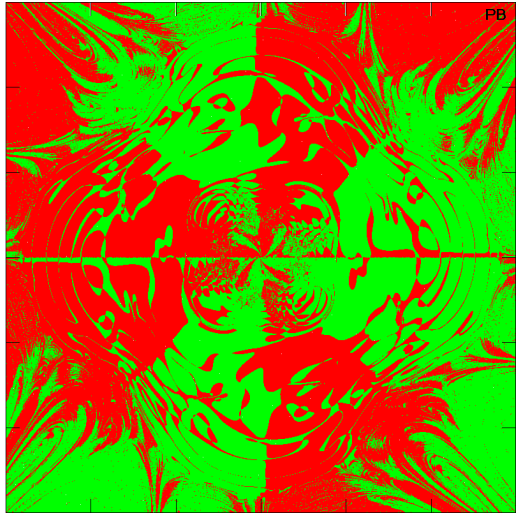}
\caption{
Red indicates a positive Lyapunov exponent in the upper half plane, negative in the lower.  If
the equations of motion are run backward by changing the righthandsides of all four equations 
the effect is to reflect the section shown here, changing the sign of $q$.  Like the other models
this Patra-Bhattacharya oscillator has a four-dimensional Gaussian distribution in its
$(q,p,\zeta,\xi)$ phase space which the dynamics samples ergodically.  Unlike the other models
about half the PB measure is regular rather than chaotic.
}
\end{figure}

The {\it four}-dimensional phase-space continuity equation shows that both the Martyna-Klein-Tuckerman
and the Hoover-Holian thermostats are consistent with exactly the same four-dimensional Gaussian,
$f_{\rm HH} = f_{\rm MKT}$ \ . Unlike the Nos\'e-Hoover single-thermostat model we believe that
the three two-thermostat models are {\it all} ergodic.  We believe that this proposition has been
thoroughly confirmed by the present work, confirming the chaos of one million independent initial
conditions in each case. In view of the lack of other suitable entries in the 2014 competition, we
have awarded ourselves the Snook Prize for this finding.  It should be noted that {\it all} of
these thermostat models, ergodic or not, {\it are} time-reversible, with the friction coefficients
$(\zeta,\xi)$ and the momentum $p$ simply changing sign if the time-ordered sequence of coordinate
values $\{ \ q(t) \ \}$ is reversed.  For this reason the time-averaged Lyapunov spectrum is
symmetric, with $\lambda_1 + \lambda_4 = \lambda_2 + \lambda_3 = \lambda_2 = \lambda_3 = 0$ .
Despite the simple analytic nature of the phase-space distribution, the chaos implied by a positive
$\lambda_1$ engenders an amazing complexity, singular in its spatial behavior, to which we turn next.

In exploring the HH, JB, and MKT  ergodic oscillator solutions we uncovered an amazingly intricate
multifractal structure most simply described in terms of the local largest Lyapunov exponent,
$\lambda_1(t)$ .  The three phase-space distributions $\{ \ f(q,p,\zeta,\xi) \ \}$ are all
multidimensional Gaussians, of the form ,
$$
f(q,p,\zeta,\xi) \propto e^{-q^2/2}e^{-p^2/2}e^{-\zeta^2/2}e^{-\xi^n/n} \ ,
$$
where $n$ is 2 for HH and MKT and 4 for JB. A closer look, using Lyapunov instability as a tool,
reveals the interesting structures shown for the three thermostated oscillators in {\bf Figures
 1-3} .  For completeness, the corresponding Patra-Bhattacharya cross section is shown as
{\bf Figure 4} .

For each of the three ergodic models we show the sign of the local Lyapunov exponent,
$\lambda_1(t)$ at the location $(q,p,0,0)$ .  The plots are analogous to a Poincar\'e section for
the more usual three-dimensional flows.  At the equilibrium, $T=1$ , these instability plots can
be generated in either of two ways: [ i ] Whenever a $(q,p,0,0)$ trajectory comes close to
$(\zeta,\xi) = (0,0)$ plot the sign of the corresponding Lyapunov exponent at the location $(q,p)$
; [ ii ] Starting at $(q,p,0,0)$ integrate {\it backwards} in time for several hundred thousand
timesteps, storing the backward trajectory. Then process the reversed trajectory data in the forward
direction, with Lyapunov instability controlling a nearby constrained satellite trajectory, finding
the local Lyapunov exponent at the chosen endpoint location $(q,p,0,0)$.  It turns out that these
two methods are roughly comparable in cost.  The second method has the advantage of providing more
detail in regions where the measure is small.  The precision shown in Figures 1-4 can be generated
in simulations
taking a day or two on a single processor.  Method [ ii ] above, reversing a stored trajectory, is
particularly well-suited to parallel simulations.  The Figures use color to indicate the sign of
the local exponent. 

\section{Fractals Lurk in the Simple Gaussian Distributions}

These cross sections have a fractal look.  In fact they are.  The reason for this was pointed out
in our paper with Dennis Isbister\cite{b13}.  As the ``past'' history of a point is being generated,
backward from the ``present'', ``bifurcations'', where the direction of the trajectory is uncertain
at the level of the numerical work limit the trustworthy scale of the Lyapunov exponent when the
stored integration is reversed, going forward in time.  For the megapixel details
shown in Figures 1-4 a few hundred thousand fourth-order ( or fifth-order or adaptive ) Runge-Kutta
timesteps suffice for reliable results at the resolution of the plotted sections.  More detail can
always be generated, at a cost exponential in the number of significant figures, by reducing the
numerical error of the trajectory integration.

The cross sections shown in the Figures have inversion symmetry.  As the source of this symmetry is
not so obvious let us describe it.  Consider a solution of the equations of motion going forward
in time,  $\{ \ q(t),p(t),\zeta(t),\xi(t),\lambda_1(t) \ \}$ \ .  Viewed simultaneously in a mirror
perpendicular to the $q$ axis an observor sees $\{ \ -q(t),-p(t),\zeta(t),\xi(t),\lambda_1(t) \ \}$ .
Evidently there is an inversion symmetry, with exactly the {\it same} Lyapunov exponent at two
different values of $(q,p)$ :
$$
\lambda_1(+q,+p) \equiv \lambda_1(-q,-p) \ .
$$
This observation saves a factor of two in computer time. The two-dimensional cross sections where
both of the friction coefficients vanish, $\zeta  = \xi = 0$ are the starting points for the backward
integrations.  The Figures indicate the magnitude of the largest Lyapunov exponent obtained at the
original starting point from a reversed trajectory, with red positive and green negative as the local
exponents varies with position. This is the case for positive $q$ .  The colors are reversed for
negative coordinate values in order to eliminate the discontinuity that would otherwise occur at
vanishing $q$ .

As is usual the fluctuations in $\lambda_1(t)$ are an order of magnitude larger than its
long-time-averaged value. The variation of the local Lyapunov exponent is not quite continuous.  If
the local exponent is plotted along a line in the $(q,p)$ plane there are occasional jumps, present
on all scales, in the exponent value.  The fractal dimension accounts for the variation of the
jump magnitudes as a function of resolution.  In the MKT case\cite{b13} the data along such a line
gave a fractal dimension of 1.69 rather than 1.00 .  Studies of the jumps along such lines reveal
their fractal structure, seen here for the first time in two dimensions.

\section{Thermostating Configurational Temperature ?}

All three of the two-thermostat HH, JB, and MKT models for heat-bath control of a harmonic oscillator
are ergodic, Because of this it was a complete surprise that Patra and Bhattacharya's control of both
the kinetic and the configurational\cite{b20} oscillator temperatures, using two thermostat variables,
was unsuccessful, leading to a mix of regular and chaotic solutions.  Their model was :
$$
\{ \ \dot q = p - \xi q\ ; \ \dot p = -q - \zeta p  \ ; \ \dot \zeta = p^2 - 1 \ ; \ \dot \xi = q^2 - 1 \ \} 
   \rightarrow \langle \ p^2, \ q^2 \ \rangle = (1, \ 1) \ [ \ {\rm PB} \ ] \ . 
$$
The stationary phase-space distribution function for the PB equations is the same as that characterizing
the HH and MKT models :
$$
f_{\rm HH} = f_{\rm MKT} = f_{\rm PB} = (2\pi)^{-2}e^{-q^2/2}e^{-p^2/2}e^{-\zeta^2/2}e^{-\xi^2/2} \ .
$$
But unlike the HH and MKT equations the Patra-Bhattacharya model {\it fails} the Lyapunov-exponent test for
ergodicity.  About half the initial conditions chosen from the four-dimensional Gaussian distribution give
regular rather than chaotic solutions.

Although controlling {\it coordinate} moments like $\langle \ q^2,q^4,q^6 \dots \ \rangle $ may
seem unphysical Landau and Lifshitz showed ( in the 1951 Russian edition of their {\it Statistical
Physics} ) that what later came to be termed a ``configurational temperature''\cite{b20}, based on a
mean-squared force, can be derived from Gibbs' canonical distribution.  A single integration by parts,
in the canonical ensemble, relates the mean-squared force to the curvature of the potential through the
temperature $T$ :
$$
kT\int [ \ \nabla^2{\cal H} \ ]e^{-{\cal H}/kT}dq \equiv \int [ \ (\nabla{\cal H})^2 \ ] e^{-{\cal H}/kT}dq \ .
$$
So control of $\langle \ q^2 \ \rangle$ can be viewed as regulating the fluctuations in the {\it forces}.
This concept is more appealing than coordinate control.  But the lack of a physical ``thermometer'' capable
of measuring force-based temperature is a disabling disadvantage of configurational temperature.  The
possibility of divergent or negative values of the instantaneous configurational temperature are further
caution flags.  The Campisi ``thermostat'' is an example of both defects :
$$
{\cal H} = (p^2/2) + (T/2)\ln (q^2) \rightarrow \{ \ \dot q = p \ ; \ \dot p = -(T/q) \ \} \ .
$$
Formally, Campisi dynamics is consistent with a Gaussian momentum distribution. Although at first
glance appealing, this simple model exhibits a {\it negative} configurational temperature, $-T$ , in one
dimension and a divergent one in two\cite{b21}.  Disagreement between the kinetic and configurational
temperatures is a clear violation of the Zeroth Law of Thermodynamics, to which we turn next.

\section{Coupled Thermostats-the Zeroth Law of Thermodynamics}

To what extent do these minimalistic thermostats, described by one or two additional control
variables, exhibit the characteristic properties of macroscopic thermal baths?  An essential
characteristic of such baths is their consistency with the Zeroth Law of Thermodynamics.  Two
``good'' thermal baths, both at a temperature $T$, should, when coupled together or to another
canonical equilibrium system at the same temperature $T$, remain at
that common temperature without modifying Gibbs' equilibrium distribution in either bath or in
another equilibrium system.
 
For our thermostated oscillators this desirable property can be tested by coupling {\it pairs}
of heat baths together with a simple Hooke's-Law spring.  The modification of the equations of
motion for oscillator number 1 and oscillator number 2, both at unit temperature, are as follows :
$$
\phi_{12} = (q_1 - q_2)^2/2 \rightarrow
\dot p_1 = \dot p_1(T=1) + q_2 - q_1 \ ; \
\dot p_2 = \dot p_2(T=1) + q_1 - q_2 \ . 
$$

It is easy to verify that the two Nos\'e-Hoover oscillators, coupled together by a spring, have
nonergodic solutions, corresponding to a six-dimensional analog of the tori making up most
of the single-oscillator phase space.  One such solution occurs with initial values of the
two oscillator momenta $(p_1,p_2)=(0.99,1.01)$ .  Accordingly we restrict our detailed investigation
to the three ergodic two-thermostat heat baths to see whether or not they can obey the Zeroth Law.

All the six combinations of like and unlike {\it pairs} of ergodic two-thermostat oscillators :
\begin{center}
[ HH+HH, HH+JB, HH+MKT, JB+JB, JB+MKT, MKT+MKT ] , \\
\end{center}
\noindent
provide chaotic eight-dimensional regions in phase space with symmetric Lyapunov spectrum.  This
symmetry implies ergodicity.  {\bf Figure 5} shows both the spectra and the spectral sums ( which
give the growth rates of phase-space hyperspheres as functions of their dimensionality ). Even
though there are relatively small but significant differences between the HH, JB, and MKT Lyapunov
exponents pairing {\it any two} of these systems together ( with a Hooke's-Law spring ) gives no
indication of the dissipation which would result if one phase-space bath were to grow at the
expense of the other. We conclude from these six examples that the HH, JB, and MKT heat baths
{\it all} obey the Zeroth Law of Thermodynamics, quite a good outcome for dynamical systems which
represent a thermodynamic heat bath with the low cost of only four phase-space dimensions.

\begin{figure}
\vspace{1 cm}
\includegraphics[height=7cm,width=6cm,angle=-90]{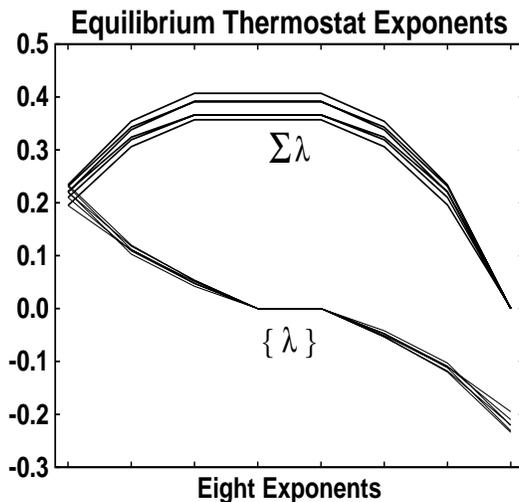}
\caption{
The eight equilibrium Lyapunov exponents describing pairs of ergodic thermostats linked by a
Hooke's-Law spring exhibit pairing when time averaged.  The JB thermostats exhibit slightly
larger Lyapunov exponents due to the quartic terms in their equations of motion.  The summed
spectra shown in the six upper lines give the growth rates of 1-, 2-, 3-, ... 8-dimensional
phase volumes.  The last sum is precisely zero as the equilibrium equations of motion conserve
phase volume when time-averaged.
}
\end{figure}

The largest Lyapunov exponent in these eight-dimensional problems varies from 0.20 for two coupled
MKT oscillators to 0.23 for two coupled JB oscillators.   The exponent for two HH oscillators lies
in between.  Evidently the quartic forces in the Ju-Bulgac and Hoover-Holian thermostats are slightly
more effective in promoting chaos than are the cubic Martyna-Klein-Tuckerman forces. This same
conclusion follows from the time required to generate the structures seen in Figures 1-4 .  The additional
chaos comes at the ( rather small ) price of an increased stiffness in the differential equations.
If any of the three ergodic thermostats were not ergodic we would expect to see a dissipative strange
attractor result in coupling it to an ergodic heat bath.  In all six pairings of the equilibrium ergodic
thermostats there is no evidence of dissipation or reduced phase-space dimensionality.  Evidently
all three $(\zeta,\xi)$ thermostats are good heat baths from the standpoint of the Zeroth Law of
Thermodynamics. Coupling pairs of thermostated oscillators with a harmonic coupling reveals that all
three of the ergodic models $\{$ HH, JB, MKT $\}$ remain ergodic when coupled and also provide the
complete canonical distribution.

\section{Nonequilibrium Simulations}

The similar equilibrium behavior of the HH, JB, and MKT thermostats suggests a further comparison for
simple heat-flow problems in which two coupled thermostats have {\it different} temperatures, leading
to hot-to-cold heat flow and dissipation.  Dissipation is reflected in phase space by the formation
of strange attractors with a fractional dimensionality reduced below the equilibrium value.  In
the linear-response regime the six two-bath possibilities must necessarily agree with Green and
Kubo's theory as all the two-constant heat baths reproduce Gibbs' canonical $(q,p)$ distribution
at equilibrium.  Let us be adventurous and treat instead the relatively far-from-equilibrium coupling
of two thermostats with thermostat 1 at half the temperature of thermostat 2 :
$$
\dot p_1 = \dot p_1(T=1) + q_2 - q_1 \ ; \
\dot p_2 = \dot p_2(T=2) + q_1 - q_2 \ .
$$
Just as before all the masses, force constants, and relaxation times are chosen equal to unity.

The far-from-equilibrium Lyapunov data corresponding to the nine combinations of thermostats,
with temperatures of 1 and 2, are shown in {\bf Figure 6} , both individually and as sums.  The
broad maximum corresponding to sums of three or four exponents indicates that the maximum growth
rate in phase space applies to four-dimensional volumes.  The summed-up spectrum crosses zero
near an abscissa value of 7 indicating that the dimensionality loss for this far-from-equilibrium
problem is on the order of 1.  Considerably larger losses, comparable to the number of particles,
occur in ``$\phi^4$-model'' chain simulations where each of a few dozen particles is tethered to
its lattice site with a cubic restoring force\cite{b7}.
\begin{figure}
\vspace{1 cm}
\includegraphics[height=7cm,width=6cm,angle=-90]{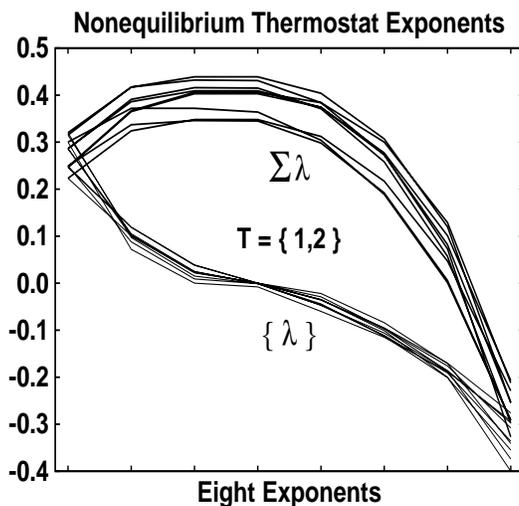}
\caption{
The eight exponents describing the coupling of ergodic thermostats with markedly different
temperatures, 1 and 2.  Spectra for all nine combinations of thermostats are shown.  The hot-to-cold
heat current dissipates phase volume.  On the average the phase volume approaches zero ( a strange
attractor ) as $\exp[ \ \sum \lambda t \ ] \simeq e^{-0.25t}$ despite the time-reversible nature of all
the equations of motion.
}
\end{figure}

Closer to equilibrium, with cold and hot temperature of 1.0 and 1.1, all nine thermostat combinations
have a Lyapunov sum of order -0.01 while temperatures of 1.0 and 2.0 provide dissipative sums in the
range from -0.20 to -0.30 .  Compare {\bf Figures 6 and 7}.  Just as in heat conduction according to
Fourier's Law, we would expect a linear-response dissipation {\it quadratic} in the temperature
difference driving the flow, varying as $(T_{\rm hot} - T_{\rm cold})^2$ .
\begin{figure}
\vspace{1 cm}
\includegraphics[height=7cm,width=6cm,angle=-90]{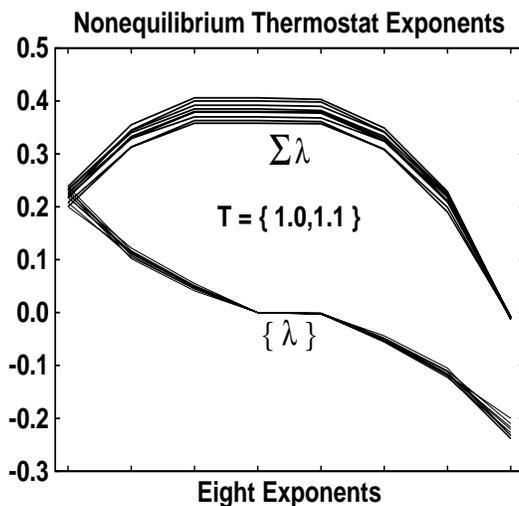}
\caption{
The eight exponents describing the coupling of ergodic thermostats with similar temperatures,
1 and 1.1.  All nine combinations are shown.  The hot-to-cold heat current could be estimated with
linear response theory and is essentially the same for all three thermostat types because the
equilibrium phase-space distribution is the same Gaussian for all three thermostat types,
$f(q,p) \propto e^{-q^2/2}e^{-p^2/2} \ $.
}
\end{figure}

Loss of phase-space volume corresponds to thermodynamic dissipation ( through Gibbs' relationship
of phase volume to entropy, $S = k\ln\Omega$, where $\Omega$ is phase volume or number of states ).
Coupling two baths, with the cooler bath first, gives the following ``entropy production rates'' :
$$
{\rm HH+ HH} : 0.322 \ ; \ {\rm HH+JB} : 0.254 \ ; \ {\rm HH+MKT} : 0.213 \ ;
$$
$$
{\rm JB+ HH} : 0.299 \ ; \ {\rm JB+ JB} : 0.253 \ ; \ {\rm JB+MKT} : 0.207 \ ;
$$ 
$$
{\rm MKT+HH} : 0.291 \ ; \ {\rm MKT+JB} : 0.290 \ ; \ {\rm MKT+MKT} : 0.229 \ .
$$
We used quote marks above to remind the reader that simply summing the spectrum is not the same as
averaging the dissipation for the cold and hot thermostats.

Dissipation can also be reckoned in terms of the phase-space dimensionality of the dissipative
strange attractor, using the Kaplan-Yorke linear interpolation between the last positive dimension
and the first negative one.  For the nine heat-bath pairings the dimensionalities are :
$$
{\rm HH+ HH} : 7.19 \ ; \ {\rm HH+JB} : 7.32 \ ; \ {\rm HH+MKT} : 7.38 \ ;
$$
$$                
{\rm JB+ HH} : 7.16 \ ; \ {\rm JB+ JB} : 7.25 \ ; \ {\rm JB+MKT} : 7.33 \ ;
$$
$$
{\rm MKT+HH} : 7.02 \ ; \ {\rm MKT+JB} : 7.01 \ ; \ {\rm MKT+MKT} : 7.17 \ .
$$
These results should provide good benchmarks for nonlinear theories of transport.

\section{Jumps -- Equilibrium Mexican Hat Potential Simulations}

\begin{figure}
\vspace{1 cm}
\includegraphics[height=7cm,width=6cm,angle=-90]{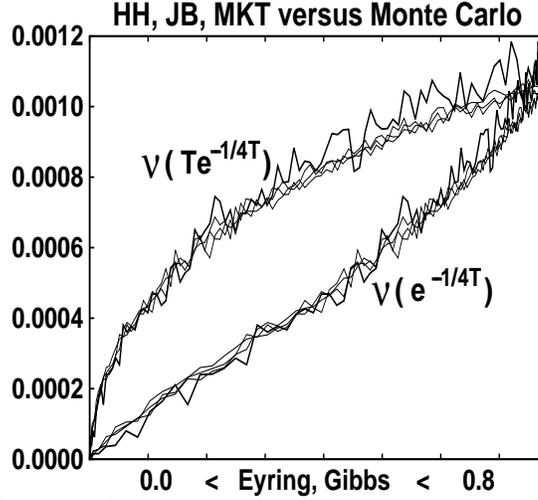}
\caption{
Coupling a Mexican Hat potential with twin well depths of 0.25 at $q = \pm 1$ to any of the
three ergodic heat baths provides a mechanism for jumping over the barrier.  The simulations
shown here indicate that all three baths provide essentially the same temperature-dependent
jump frequency $\nu$ .  The dependence is nearly linear when the coupling is small ( $\kappa = 1/10$ )
and the abscissa is Gibbs': $e^{-1/4T}$ .  Eyring's description ( the upper three curves ) is
shown where the abscissa is now $Te^{-1/4T}$ .  Gibbs' better approximates an Arrhenius straight
line.  Monte Carlo simulation of Mexican Hat dynamics, with a maximum jump length of $0.009\sqrt{T}$
, results in the heavier line plotted in addition to the results for the three heat-bath models.
}
\end{figure}

The double-well ``Mexican Hat'' potential\cite{b8} ,
$$
\phi (q) = -\left[ \ \frac{q^2}{2} \ \right]  + \left[ \ \frac{q^4}{4} \ \right] \longrightarrow
 \phi(\pm 1) = - \left[ \ \frac{1}{4} \ \right] \ ; \ \phi(0) = 0 \ ,
$$
has a barrier between its two minima of height $(1/4)$ .  At a temperature of unity, where the
barrier is negligible, simulations of the Hat potential coupled to the HH, MKT, and JB thermostats
result in excellent agreement with Gibbs' canonical distribution.  At sufficiently low temperatures
we expect that the frequency of successful jumps over the barrier will decline, as would also be
the case using the Metropolis, Rosenbluths, and Tellers' Monte Carlo scheme with a reasonable jump
length, perhaps of order 0.01 .

A straightforward ``weak coupling'' of the two-minimum Hat potential to any one of the ergodic oscillator
heat baths ,
$$
\Phi(q_{\rm Bath},q_{\rm Hat}) = \frac{1}{20}( \ q_{\rm Bath} - q_{\rm Hat} \ )^2 \ ,
$$
leads to ``jump'' frequencies that vary roughly as Gibbs' canonical barrier weight, $e^{-1/4T}$ .
Eyring's model, as mentioned in the Wikipedia article on ``Henry Eyring ( chemist )'', and with
a jump frequency varying as $Te^{-1/4T}$ , is a relatively poor description
and corresponds to the upper curves in the {\bf Figure 8} . All three ergodic thermostat models
behave similarly, as is shown in triplets of narrower lines in {\bf Figure 8} .

Another possibility for modelling jumps is to use the Metropolis, Rosenbluths, and Tellers' Monte
Carlo method. By adusting the Monte Carlo trial steplength the two approaches, dynamical and Monte
Carlo, can be made to correspond.  We have compared the number of jumps over the barrier in
billion-jump simulations with maximum jump length $dq = 0.009\sqrt{T}$ to dynamical simulations with
timestep 0.005, where in all these latter dynamical cases the Hat potential is coupled to an HH,
MKT, or JB thermostat with temperatures varying from 0.010 to 1.00 , as shown in Figure 8 .

\section{Summary}

The Lyapunov spectra for three varieties of time-reversible deterministic oscillator thermostats
show that all three of them reproduce an extended ( four-variable ) form of Gibbs' canonical
( two-variable ) distribution ,
$$
f(q,p)=e^{-q^2/2}e^{-p^2/2}/(2\pi) \ .
$$
The HH and MKT equations have Gaussian distributions for $\zeta$ and $\xi$ while the JB distribution
is Gaussian in $\zeta^2$ and $\xi$.  All three approaches are sufficiently robust to thermostat a
harmonic oscillator.  The three are also consistent with the Zeroth Law of Thermodynamics in the
sense that
any pair of them at a common temperature $T$ generates Gibbs' canonical distribution for that
temperature for both thermostats.

This set of ergodic thermostats represents a good toolkit for simulations with a few degrees of
freedom as it gives an idea of the dependence of dissipation on the chosen form of the thermostat.
With many degrees of freedom, where ergodicity is not an issue [ Poincar\'e recurrence takes
forever so that the size of fluctuations determines the usefulness of the results ] any of these
thermostats, as well as the single thermostat Nos\'e-Hoover model, are likely equally useful.

We have considered two {\it nonergodic} time-reversible thermostats, the Patra-Bhattacharya and
Nos\'e-Hoover, which reproduce the specified kinetic temperature but do not reproduce all of
Gibbs' canonical distribution.  When the PB or NH thermostats are coupled to the HH, JB, or MKT
thermostats two qualitatively different results occur: [ 1 ] a dissipative strange attractor
in the case of the PB model and [ 2 ] a conservative integer-dimensional chaotic sea when
coupled to the Hamiltonian-based Nos\'e-Hoover oscillator.  These results are similar for each
of the three ergodic thermostats.
In the PB case the dissipation is an indicator of the nonHamiltonian nature of the dynamics.
Any of the ergodic thermostats, Hoover-Holian, Ju-Bulgac, Martyna-Klein-Tuckerman, can be used
successfully in both equilibrium and nonequilibrium simulations.

The first two of these can also be applied in nonequilibrium simulations where it is desired to
specify the kinetic temperature of selected degrees of freedom.  Though there is no problem in
solving the MKT equations for nonequilibrium problems Brad Holian pointed out long ago\cite{b18}
that this thermostat fails to return its target temperature due to the nonzero correlation
linking the two MKT friction coefficients,
$\langle \ \zeta\xi \ \rangle_{\rm MKT}^{\rm noneq} \neq 0$ .

This relatively thorough investigation of ergodicity, based on one million independent initial
conditions, should settle ( at least from the numerical, as opposed to analytical, viewpoint )
the question of the ergodicity of the Martyna-Klein-Tuckerman oscillator.  The apparent
``holes'' in that oscillator's cross section are still a small ``puzzle''.  A second such
``puzzle'' is the nature of the local Lyapunov exponents.  Their intricate multifractal Lyapunov
structure is well concealed within an innocent Gaussian distribution.

From the standpoint of dynamical-systems theory ( as opposed to thermodynamics ) we wish to
emphasize the amazing nature of the chaos hidden within the simple ergodic Gaussian distributions.
The smoothness of the distributions and the good convergence of their moments conceals the fractal
nature of their chaos, as shown in Figure 1 through 3 .  We recommend all three systems for further
studies. The $(\zeta = \xi = 0)$ plane is only a single choice among the many possible.  Despite their
fractal nature the cross-sectional values are well-behaved, accessible, and reproducible from the
numerical standpoint.  It is encouraging for the future that work originally chosen to settle some
ergodicity questions turned out to generate a swarm of other problems still needing more detailed
understanding.

\section{Acknowledgments} We thank Carol Hoover, Aurel Bulgac, and Baidurya Bhattacharya, for
their able help.

\end{document}